\documentclass[journal, 10pt]{IEEEtran}

\usepackage{multirow}
\usepackage[font=scriptsize,caption=false,labelsep=space]{subfig}
\usepackage{mathrsfs}
\usepackage{graphicx,float,wrapfig,epstopdf,amsmath}
\usepackage[]{algorithmicx}
\usepackage{algpseudocode,algorithm}
\usepackage{balance}
\usepackage{amsmath}
\usepackage{amssymb}
\usepackage{amsthm}
\usepackage{graphicx}
\usepackage{epstopdf}
\usepackage{times}
\usepackage{textcomp,cite}
\usepackage{xcolor}
\usepackage{url}

\usepackage{multirow}
\usepackage{threeparttable}
\graphicspath{{./figures/}}

\newcommand{\subparagraph}{}
\usepackage[compact]{titlesec}
\titlespacing{\section}{0pt}{2ex}{1ex}
\titlespacing{\subsection}{0pt}{1ex}{0ex}
\titlespacing{\subsubsection}{0pt}{0.5ex}{0ex}

\begin{document}
	
	\title{Federated Learning for Hybrid Beamforming in mm-Wave Massive MIMO
	}
	\author{Ahmet~M.~Elbir\textit{, Senior Member, IEEE}, and Sinem Coleri\textit{, Senior Member, IEEE} 
		
		\thanks{Sinem Coleri acknowledges the support of the Scientific and Technological Research Council of Turkey for EU CHIST-ERA grant 119E350 and Ford Otosan.}
		
		\thanks{Ahmet M. Elbir is with the Department of Electrical and Electronics Engineering, Duzce University, Duzce, Turkey (e-mail: ahmetmelbir@gmail.com).} 
		\thanks{Sinem Coleri is with the Department of Electrical and Electronics Engineering, 
			Koc University, Istanbul, Turkey (e-mail: scoleri@ku.edu.tr).}
	}
	\maketitle
	
	\begin{abstract}
		Machine learning for hybrid beamforming  has been extensively studied  by using centralized machine learning (CML) techniques, which require the training of a global model with a large dataset collected from the users. However, the transmission of the whole dataset between the users and the base station (BS) is computationally prohibitive due to limited communication bandwidth and  privacy concerns. In this work, we introduce a federated learning (FL) based framework for hybrid beamforming, where the model training is performed at the BS by collecting only the gradients from the users. We design a convolutional neural network, in which the input is the channel data, yielding the analog beamformers at the output. Via numerical simulations, FL is demonstrated to be  more tolerant to the imperfections and corruptions in the channel data as well as having less transmission overhead than CML.
	\end{abstract}

	\section{Introduction}
	\label{sec:Introduciton}

	In millimeter wave (mm-Wave) massive MIMO (multiple-input multiple-output) systems, the users estimate the downlink channel and corresponding hybrid (analog and digital) beamformers, then send these estimated values to the base station (BS) via uplink channel with limited feedback techniques~\cite{mimoHybridLeus3}. Several hybrid beamforming approaches are proposed in the literature~\cite{mimoHybridLeus3,mimoRHeath}. However, the performance of these approaches strongly relies on the perfectness of the channel state information (CSI). To provide a robust beamforming performance, data-driven approaches, such as machine learning, are proposed~\cite{fastDL_HB,elbirDL_COMML,elbirHybrid_multiuser}. ML is advantageous in extracting features from the channel data and exhibits robust performance against the imperfections/corruptions at the input. 
	
\begin{figure}[h]
		\centering
		{\includegraphics[draft=false,width=\columnwidth]{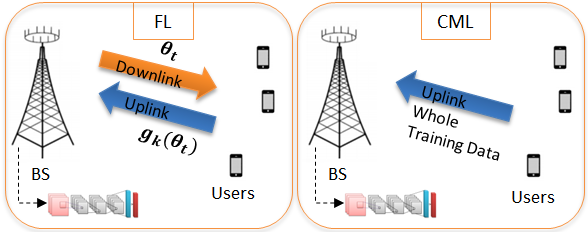} }
		\caption{Data and model transmission schemes for multi-user massive MIMO.
		}
		\label{fig_Diagram}
	\end{figure}

	In the context of ML, {\color{black}usually centralized ML (CML) techniques have been used, where }a neural network (NN) model at the BS is trained with the training data collected by the users (see Fig.~\ref{fig_Diagram}). This process requires the transmission of each user's data to the BS.
	Therefore, the data collection and transmission introduce significant overhead. To alleviate the transmission overhead problem, federated learning (FL) strategies are recently introduced, especially for edge computing applications~\cite{FL_Bennis}. In FL, instead of sending the whole data to the BS, the edge devices (mobile users) only transmit the gradient information of the NN model. Then, the NN is iteratively trained by using stochastic gradient descent (SGD) algorithm~\cite{FL_QSGD}.   Although, this approach introduces the complexity for computation of the gradients at the edge devices, transmitting only the gradient information significantly lowers the overhead of data collection and transmission.
	
	
	Up to now, FL has been mostly considered for the applications of  wireless sensor networks, e.g., UAV (unmanned aerial vehicle) networks~\cite{FL_Bennis2,FL_Bennis3}, vehicular networks~\cite{FL_Bennis5,elbir2020federated}. In \cite{FL_Bennis2}, the authors applies FL to the trajectory planning of a UAV  swarm and the data collected by  each UAV are processed for gradient computation and a global NN at the ``leading" UAV is trained. In \cite{FL_Bennis3} and \cite{FL_Bennis5}, client scheduling and power allocation problems are investigated for FL framework, respectively. Furthermore, a review of FL for vehicular network is presented in~\cite{elbir2020federated}. {\color{black}In addition, FL is considered in \cite{FL_traffic} for the classification of encrypted mobile traffic data.} Different from~\cite{FL_Bennis2,FL_Bennis3,FL_Bennis5,elbir2020federated,FL_traffic}, \cite{FL_Gunduz} considers a more realistic scenario where the gradients are transmitted to the BS through a noisy wireless channel and a classification model is trained for image classification. Notably, to the best of our knowledge, the application of FL to massive MIMO hybrid beamforming is not investigated in the literature.  Furthermore, above works mostly consider simple {\color{black}CML} structures, which accommodate shallow NN architectures, such as a single layer NN~\cite{FL_Gunduz}. The performance of these architectures can be leveraged by utilizing deeper NNs, such as convolutional neural networks (CNNs)~\cite{elbirQuantizedCNN2019,elbirHybrid_multiuser}, which also motivates us to exploit {\color{black}CNN architectures} for FL in this work.

	{\color{black}In this letter, we propose a federated learning (FL) framework for hybrid beamforming.} We design a CNN architecture employed at the BS. The CNN accepts channel matrix as input and yields the RF beamformer at the output. The deep network is then trained by using the gradient data collected from the users. Each user computes the gradient information with its available training data (a pair of channel matrix and corresponding beamformer index), then sends it to the BS. The BS receives all the gradient data from the users and performs parameter update for the CNN model. Via numerical simulations, the proposed FL approach is demonstrated to provide robust beamforming performance while exhibiting much less {\color{black}transmission overhead}, compared to {\color{black}CML} works~\cite{elbirHybrid_multiuser,fastDL_HB,DL_HB_multiuser}.
	
	{\color{black}In the remaining of this work, we introduce the signal model in Sec.~\ref{sec:SystemModel} and present the proposed FL approach for hybrid beamforming in Sec.~\ref{sec:FL4HB}. After presenting the numerical results for performance evaluation in Sec.~\ref{sec:Sim}, we conclude the letter in Sec.~\ref{sec:Conc}.     }

	\section{Signal Model and Problem Description}
	\label{sec:SystemModel}
	We consider a multi-user MIMO scenario where the BS, having $N_\mathrm{T}$ antennas, communicates with $K$ single-antenna users.	In the downlink, the BS first precodes $K$ data symbols $\mathbf{s} = [s_1,s_2,\dots,s_K]^\textsf{T}\in \mathbb{C}^{K}$  by applying  baseband precoders $\mathbf{F}_{\mathrm{BB}} = [\mathbf{f}_{\mathrm{BB},1},\dots,\mathbf{f}_{\mathrm{BB},{K}}]\in \mathbb{C}^{K\times K}$, and then,  employs $K$ RF precoders $\mathbf{F}_{\mathrm{RF}} = [\mathbf{f}_{\mathrm{RF},1},\dots, \mathbf{f}_{\mathrm{RF},K}]\in \mathbb{C}^{N_\mathrm{T}\times K}$ to form the transmitted signal. Given that $\mathbf{F}_{\mathrm{RF}}$ consists of analog phase shifters, we assume that the RF precoder has constant unit-modulus elements, i.e., $|[\mathbf{F}_{\mathrm{RF}}]_{i,j}|^2 =1$, thus,  $ \|\mathbf{F}_{\mathrm{RF}}\mathbf{F}_{\mathrm{BB}} \|_\mathcal{F}^2= K$. Then, the $N_\mathrm{T}\times 1$ transmit signal is $	\mathbf{x} = \mathbf{F}_{\mathrm{RF}} \mathbf{F}_{\mathrm{BB}}  \mathbf{s},$
	and the received signal at the $k$-th user becomes
	$	{y}_k = \mathbf{h}_k^\textsf{H} \sum_{n=1}^{K}\mathbf{F}_{\mathrm{RF}} \mathbf{f}_{\mathrm{BB},n}  {s}_n + {n}_k,$
	where $\mathbf{h}_k\in \mathbb{C}^{N_\mathrm{T}}$ denotes the mm-Wave channel for the $k$-th user with $\|\mathbf{h}_k \|_2^2=N_\mathrm{T}$ and ${n}_k \sim \mathcal{CN}({0}, \sigma^2)$ is the additive white Gaussian noise (AWGN) vector. 
	
	
	We adopted clustered channel model where the mm-Wave channel $\mathbf{h}_k$ is represented as the cluster of $L$ line-of-sight (LOS) received path rays~\cite{mimoRHeath,mimoHybridLeus3}, i.e.,
	\begin{align}
	\mathbf{h}_k = \beta \sum_{l} \alpha_{k,l} 	\mathbf{a} (\varphi^{(k,l)}),
	\end{align}
	where $\beta = \sqrt{N_\mathrm{T}/L }$, $\alpha_{k,l}$ is the complex channel gain, $\varphi^{(k,l)}\in [\frac{-\pi}{2},\frac{\pi}{2}]$ denotes the direction of the transmitted paths, $\mathbf{a}(\varphi^{(k,l)}) $  is the ${N}_\mathrm{T}\times 1$ array steering vector whose $m$-th element is given by $	\big[ \mathbf{a}(\varphi^{(k,l)}) \big]_m = \exp \{-j\frac{2\pi }{\lambda } d(m-1)\sin(\varphi^{(k,l)})   \},$
	where $\lambda$ is the wavelength and $d$ is the array element spacing.
	
	Assuming Gaussian signaling, then the achievable rate for the $k$-th user is $R_k = \log_2 \bigg| 1 +  \frac{ \frac{1}{K}| \mathbf{h}_k^\textsf{H} \mathbf{F}_\mathrm{RF} \mathbf{f}_{\mathrm{BB}_k}   |^2    }{ \frac{1}{K} \sum_{n \neq k}| \mathbf{h}_n^\textsf{H} \mathbf{F}_\mathrm{RF} \mathbf{f}_{\mathrm{BB}_n}   |^2 + \sigma^2  } \bigg|$ and the sum-rate is $\bar{R} = \sum_{k\in \mathcal{K}} R_k$ for $\mathcal{K} = \{1,\dots,K\}$~\cite{mimoHybridLeus3}. 
	
	Let us denote the parameter set of the global model at the BS as $\boldsymbol{\theta}\in \mathbb{R}^P$, which has $P$ real-valued learnable parameters. Then, the learning model can be represented by the nonlinear relationship between the input $\mathcal{X}$ and output $\mathcal{Y}$, given by $	f (\boldsymbol{\theta}, \mathcal{X} ) = \mathcal{Y}.$
	
	In this work, we focus on the training stage of the global NN model. {\color{black}Therefore, we assume that each user has available training data pairs, e.g., the channel vector $\mathbf{h}_k$ (input) and the corresponding RF precoder $\mathbf{f}_{\mathrm{RF},k}$ (output label).  It is worthwhile to note that channel and beamformer estimation can be performed by both ML~\cite{elbirHybrid_multiuser,elbirQuantizedCNN2019,elbir2019online} and non-ML~\cite{mimoRHeath,mimoHybridLeus3} algorithms.} The aim in this work is to learn $\boldsymbol{\theta}$ by training the global NN with the available training data available at the users. Once the learning stage is completed, the users can predict their corresponding RF precoder by feeding the NN with the channel data, then feed it back to the BS.

	\section{FL For Hybrid Beamforming}
	\label{sec:FL4HB}
	Let us denote the training dataset for the $k$-th user by $\mathcal{D}_k = \{(\mathcal{X}_k^{(1)},\mathcal{Y}_k^{(1)}),\dots,(\mathcal{X}_k^{(\textsf{D}_k)},\mathcal{Y}_k^{(\textsf{D}_k)}) \}$, where $\textsf{D}_k = |\mathcal{D}_k|$ is the size of the dataset and $\mathcal{X} = \bigcup_{k\in \mathcal{K}} \mathcal{X}_k$, $\mathcal{Y} = \bigcup_{k\in \mathcal{K}} \mathcal{Y}_k$. In {\color{black}CML}-based works~\cite{fastDL_HB,elbirHybrid_multiuser,elbirQuantizedCNN2019,elbirDL_COMML}, the training of the global model is performed at the BS by collecting the datasets of all users (see Fig.~\ref{fig_Diagram}). Once the BS collects $\mathcal{D} = \bigcup_{k=1}^K \mathcal{D}_k$, the global model is trained by minimizing the empirical loss,
	\begin{align}
	\label{lossML}
	\mathcal{F}(\boldsymbol{\theta}) =  \frac{1}{\textsf{D}} \sum_{i = 1}^{\textsf{D}}\mathcal{L}(f(\boldsymbol{\theta}; \mathcal{X}^{(i)}),\mathcal{Y}^{(i)}  ),
	\end{align}
	where $\textsf{D} = |\mathcal{D}|$ is the size of the training dataset and $\mathcal{L}(\cdot)$ is the loss function defined by the learning model. The minimization of empirical loss is achieved via gradient descent (GD) by updating the model parameters $\boldsymbol{\theta}_t$ at iteration $t$ as
	$\boldsymbol{\theta}_{t+1} = \boldsymbol{\theta}_t - \eta_t \nabla \mathcal{F}(\boldsymbol{\theta}_t),$
	where $\eta_t$ is the learning rate at $t$. However, for large datasets, the implementation of GD is computationally prohibitive. To reduce the complexity, SGD technique is used, where $\boldsymbol{\theta}_t$ is updated by
	\begin{align}
	\boldsymbol{\theta}_{t+1} = \boldsymbol{\theta}_t - \eta_t \mathbf{g}(\boldsymbol{\theta}_t),
	\end{align}
	which satisfies $\mathbb{E}\{ \mathbf{g}(\boldsymbol{\theta}_t) \} = \nabla \mathcal{F}(\boldsymbol{\theta}_t)$. Therefore, SDG allows us to minimize the empirical loss by partitioning the dataset into multiple portions. In {\color{black}CML}, SGD is mainly used to accelerate the learning process by partitioning the training dataset into batches, which is known as mini-batch learning~\cite{FL_QSGD}. On the other hand, in FL, the training dataset is partitioned into small portions, however they are available at the edge devices. Hence, $\boldsymbol{\theta}_t$ is  updated, by collecting the local gradients $\{\mathbf{g}_k(\boldsymbol{\theta}_t)\}_{k\in \mathcal{K}}$  computed at users with their own datasets  $\{\mathcal{D}_k\}_{k\in \mathcal{K}}$. Thus, the BS incorporates $\{\mathbf{g}_k(\boldsymbol{\theta}_t)\}_{k\in \mathcal{K}}$ to update the global model parameters as
	\begin{align}
	\boldsymbol{\theta}_{t+1} = \boldsymbol{\theta}_t - \eta_t  \frac{1}{K} \sum_{k=1}^{K} \mathbf{g}_k(\boldsymbol{\theta}_t),
	\end{align}
	where $\mathbf{g}_k(\boldsymbol{\theta}_t) = \frac{1}{\textsf{D}_k} \sum_{i=1}^{\textsf{D}_k}\nabla \mathcal{L} (f (\boldsymbol{\theta}_t; \mathcal{X}_k^{(i)}, \mathcal{Y}_k^{(i)})  )   $ is the stochastic gradient computed at the $k$-th user with $\mathcal{D}_k$.

	Due to the limited number of users, $K$, there will be deviations in the gradient average  $\frac{1}{K}\sum_{k=1}^K\mathbf{g}_k(\boldsymbol{\theta}_t)$ from the stochastic average $\mathbb{E}\{ \mathbf{g}(\boldsymbol{\theta}_t)\}$. To reduce the oscillations due to the gradient averaging, parameter update is performed by using a momentum parameter $\gamma$ which allows us to ``moving-average'' the gradients~\cite{FL_QSGD}. Finally, the parameter update with momentum is given by
	\begin{align}
	\label{gradientUpdateWithmomentum}
	\boldsymbol{\theta}_{t+1} = \boldsymbol{\theta}_t - \eta_t  \frac{1}{K} \sum_{k=1}^{K} \mathbf{g}_k(\boldsymbol{\theta}_t) + \gamma (\boldsymbol{\theta}_t - \boldsymbol{\theta}_{t-1}).
	\end{align}

	Next, we discuss the input-output data acquisition and network architecture for the proposed FL framework.

	\begin{figure*}[h]
		\centering
		\subfloat[]{\includegraphics[draft=false,width=\columnwidth]{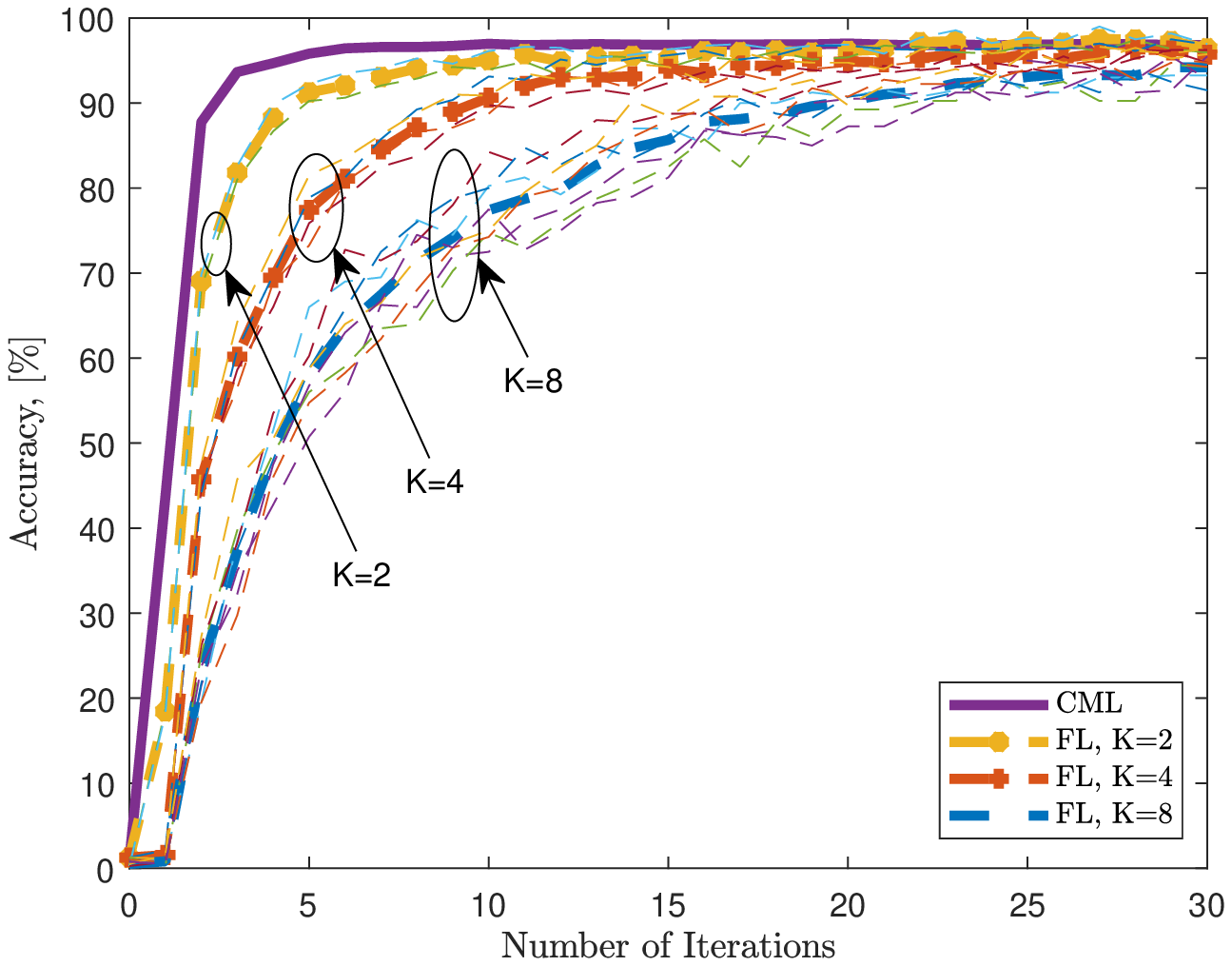} } 
		\subfloat[]{\includegraphics[draft=false,width=\columnwidth]{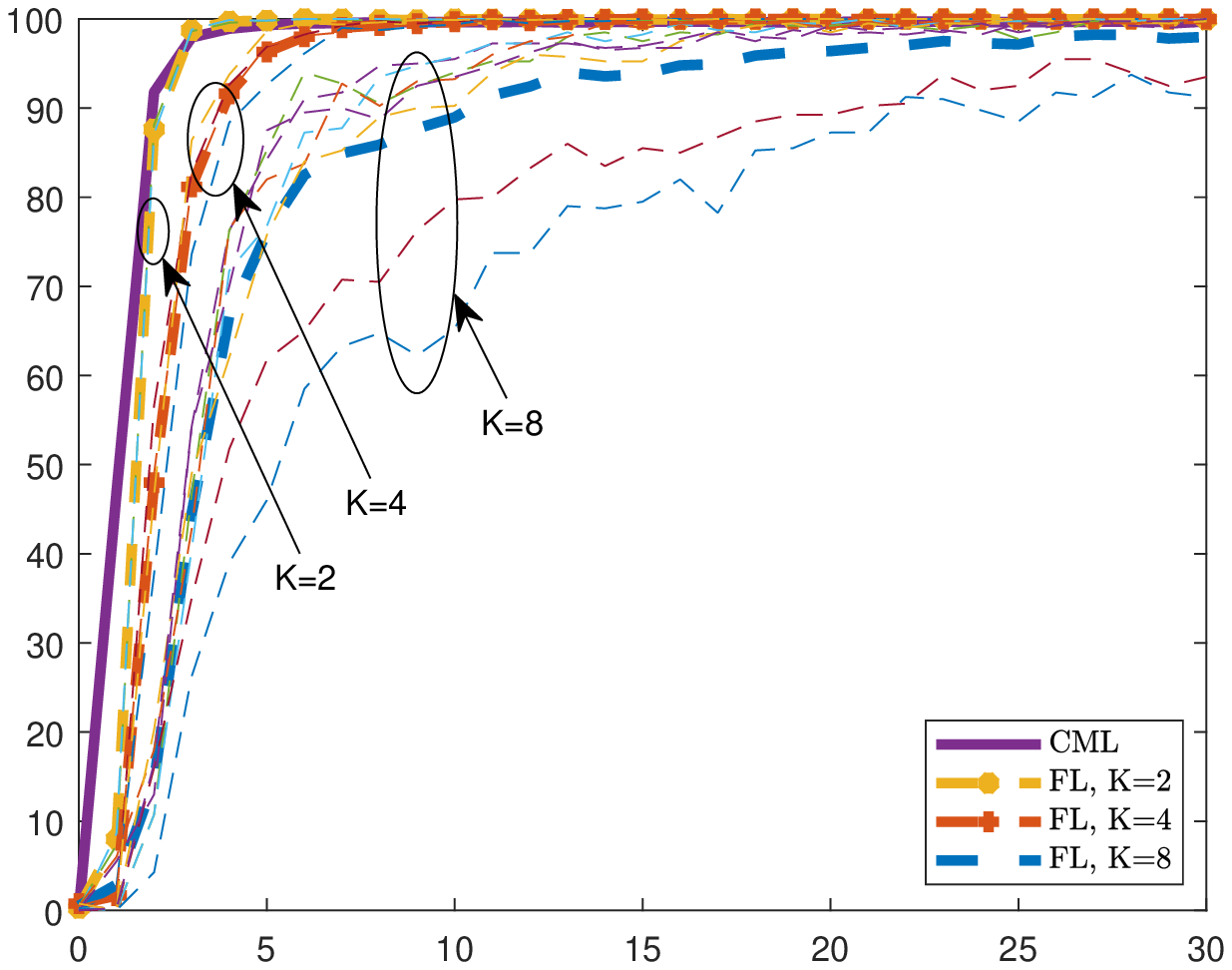} } 
		\caption{\color{black}Validation accuracy for different number of users for (a) \emph{scenario 1}  and (b) \emph{scenario 2,} respectively.
		}
		\label{fig_K_test}
	\end{figure*}
	
	\subsection{Data Acquisition}
	In this part, we discuss how the acquisition is done for training process. The input of the NN is a real-valued three ``channel'' tensor, i.e., $\mathcal{X}_k\in \mathbb{R}^{\sqrt{N_\mathrm{T}}\times \sqrt{N_\mathrm{T}}\times 3}$. {\color{black}The reason for using ``three-channel'' input is to improve the input feature representation by using the feature augmentation method~\cite{elbirDL_COMML,elbirHybrid_multiuser}}. We first reshape the channel vector by using a function $\Pi(\cdot):\mathbb{R}^{N_\mathrm{T}}\rightarrow \mathbb{R}^{\sqrt{N_\mathrm{T}}\times \sqrt{N_\mathrm{T}}}$ as $	\mathbf{H}_k = \Pi(\mathbf{h}_k),$
	which concatenates the $\sqrt{N_\mathrm{T}}\times 1$ sub-columns of $\mathbf{h}_k$ into a matrix. We use 2-D input data since it provides better feature representation {\color{black} and extraction for 2-D }convolutional layers. Also, we assume that $\sqrt{N_\mathrm{T}}$ is an integer value, if not, $\mathbf{H}_k$ can always be constructed as a rectangular matrix. Then, for the $k$-th user, we construct the first and the second ``channel" of input data as $[\mathcal{X}_k]_1 = \operatorname{Re}\{\mathbf{H}_k \}$ and  $[\mathcal{X}_k]_2 = \operatorname{Im}\{\mathbf{H}_k \}$, respectively. Also, we select the third ``channel'' as element-wise phase value of $\mathbf{H}_k$ as $[\mathcal{X}_k]_3 = \angle\{ \mathbf{H}_k \}$.
	
	The output layer of the NN is a classification layer which determines the class/index assigned to each RF precoder corresponding to the user directions.  Hence, we first divide the angular domain $\Theta = [\frac{-\pi}{2}, \frac{\pi}{2}]$ into $Q$ equally-spaced non-overlapping subregions $\Theta_q=[\varphi_q^\mathrm{start},\varphi_q^\mathrm{end} ]$ for $q \in \mathcal{Q} = 1,\dots,Q$ such that  $\Theta = \bigcup_{q\in \mathcal{Q}}\Theta_q$. These subregions represent the $Q$ classes used to label the whole dataset. For example, let us assume that the $k$-th user is located in $\Theta_{q}$. Then  we represent the channel data $\mathbf{h}_k$ is labeled by $q$, i.e., $\mathcal{Y}_k=q$. In other words, $\mathbf{f}_{\mathrm{RF},k}$ is constructed as $\mathbf{a}(\tilde{\varphi}_q)$ where $\tilde{\varphi}_q = \frac{\varphi_q^\mathrm{start}+\varphi_q^\mathrm{end}}{2}$.
	
	{\color{black}During training, the gradient data are received at the BS in synchronous fashion, i.e., the users' gradient information (e.g., $\mathbf{g}_k(\boldsymbol{\theta}_t)$) are aligned through a synchronization mechanism and they are received in the same  state $\boldsymbol{\theta}_{t+1}$ after model aggregation~\cite{FL_Gunduz,FL_traffic}. The model aggregation is performed with model update frequency that controls the duration of the training stage. We assume that the model update is repeated periodically with period set to channel coherence interval, in which the gradients of users are collected at the BS~\cite{fl_singleCoherenceInterval}. }

	After training, each user has access to the learned global model $\boldsymbol{\theta}$. Hence, they simply feed the NN with $\mathbf{h}_k$ to obtain the output label $q$. Then, the analog beamformer can be constructed as the steering vector $\mathbf{a}(\tilde{\varphi}_q)$ with respect to the direction $\tilde{\varphi}_q$ as defined above. Once, the BS has $\mathbf{F}_\mathrm{RF}$, the baseband beamformer $\mathbf{F}_\mathrm{BB}$ can be designed accordingly to suppress the user interference~\cite{mimoHybridLeus3}.

	\subsection{Deep Network Architecture}
	The global model is comprised of $11$ layers with two {\color{black}2-D} convolutional layers  (CLs) and a single fully connected layer (FCL). The first layer is the input layer of size $\sqrt{N_\mathrm{T}}\times \sqrt{N_\mathrm{T}}\times 3$. The second and the fifth layers are CLs, each of which has $N_\mathrm{CL}=256$ filters of size $3\times 3$. The eighth layer is an FCL which has $N_\mathrm{FCL}=512$ units. After each CL, there is a normalization layer following a $\mathrm{ReLU}$ layer, i.e., $\mathrm{ReLU}(x) = \max (0,x)$. The ninth layer is a dropout layer with {\color{black}$\zeta = 50\%$} probability factor after the FCL. The tenth layer is a  $\mathrm{softmax}$ layer. Finally,  the output layer is the classification layer of size $Q$. The proposed CNN has {\color{black} $P= \underbrace{2(CN_\mathrm{CL} W_x W_y)}_{\mathrm{CL}} +   \underbrace{\zeta N_\mathrm{CL}  W_x W_yN_\mathrm{FCL} \footnotesize }_{\mathrm{FCL}}$ parameters.} Here, $C=3$ is the number of ``channels" and $W_x =W_y=3$ are the 2-D size of each filter. Consequently, we have $P=13824 + 589824 = 603648$ and  observe that CLs provide more efficient way of reducing the network size than FCL.

	\section{Numerical Simulations}
	\label{sec:Sim}
	In this section, we evaluate the performance of the proposed {\color{black}FL for hybrid beamforming, henceforth called FLHB,} approach via numerical simulations. First, we compare {\color{black}FLHB} with {\color{black}CML} for learning performance. Then, we compare the beamforming performance of {\color{black}FLHB} with both {\color{black}CML}-based MLP (multilayer perceptron)~\cite{fastDL_HB}  and model-based SOMP (spatial orthogonal matching pursuit)~\cite{mimoRHeath}  techniques.

	The proposed FL model is realized and trained in MATLAB on a PC with a single GPU and a 768-core processor. We use the SGD algorithm with momentum $\gamma=0.9$  and  updated the network parameters with learning rate $0.001$. {\color{black}The mini-batch size for CML training is $256$.}
	
	For training dataset, we generate $N=500$ different channel realizations for each user, which are randomly located in $\Theta$ with $L=5$ paths and angle spread of $\sigma_{\varphi} = 3^\circ$. {\color{black}To improve the robustness of the NN, we also add synthetic AWGN onto each channel realization to generate $G=100$ noisy channel scenarios.}	{\color{black}In order to model the user locations realistically, we consider two scenarios. \emph{Scenario 1:} The user directions are uniformly distributed, i.e.,  $\varphi^{(k,l)}\sim \mathrm{unif}(\Theta) $, $\forall k,l$  with probability $P(\varphi^{(k,l)} \in \Theta) = 1/Q$. \emph{Scenario 2:} $\Theta$ is divided into $K$ non-overlapping equally-spaced subregions, i.e., $\{\tilde{\Theta}_k\}_{k\in \mathcal{K}}$. Then, we assume that the $k$-th user is located in the subregion $\tilde{\Theta}_k$, such that $\Theta = \bigcup_{k\in \mathcal{K}}\tilde{\Theta}_k$. Hence, the latter provides a more realistic way of representing the user locations while the former is the same as i.i.d. dataset.} For {\color{black}FLHB} and {\color{black}CML}, training is performed for the same NN with the same training dataset.  We present the results for different number of users while keeping the dataset size fixed as $\textsf{D}= 3\cdot N\cdot G \cdot K = 1,200,000$ by selecting $G = 100\cdot \frac{8}{K}$. The only difference for {\color{black}FLHB} is that, the training data is partitioned into $K$ blocks and the weight update is done by using the averaged gradients as in (\ref{gradientUpdateWithmomentum}).

	We assume that the BS has $N_\mathrm{T}=100$ antennas with $d=\frac{\lambda}{2}$ and there are $K=8$ users, unless stated otherwise. We add synthetic AWGN into each channel data for $G=100$ realizations with respect to $\mathrm{SNR}_\mathrm{TRAIN} =  20\log_{10}(\frac{|[\mathbf{H}_k]_{i,j}|^2}{\sigma_{\mathbf{H}}^2})$  to reflect the imperfect channel conditions and select $\mathrm{SNR}_\mathrm{TRAIN}=\{15, 20, 25\}$ dB. {\color{black}Hence, the size of the $4$-D training data is  $\sqrt{N_\mathrm{T}}\times \sqrt{N_\mathrm{T}}\times 3\times \textsf{D} = 10\times 10\times 3\times 1,200,000$, and we select the number of classes as $Q=360$.}
	{\color{black}During training, both $k$-fold cross-validation (e.g., $k=5$) and hold-out (e.g., $5$ to $1$ separation) techniques are employed and analogous results are obtained due to the usage of large dataset. Hence, we use the hold-out approach in the final settings by randomly separating the whole dataset as $80\%$ and $20\%$ for training and validation, respectively. When the dataset is relatively small, then $k$-fold cross-validation~\cite{FL_traffic} and transfer learning~\cite{elbir2020TL} techniques can be used to obtain a well-trained NN.} 	Once the training is completed, validation data is used in prediction stage. However, we add a synthetic noise into the validation data by $\mathrm{SNR}_\mathrm{TEST}$ to represent the imperfect channel data.

	%
	%
	%

	Fig.~\ref{fig_K_test} shows the training performance of {\color{black}FLHB} and {\color{black}CML} for $K=\{2, 4, 8\}$ in \emph{scenario 1} (Fig.~\ref{fig_K_test}(a)) and \emph{scenario 2} (Fig.~\ref{fig_K_test}(b)), respectively. The accuracy of each user's dataset $\textsf{D}_k$ is also shown in dashed curves to demonstrate the deviations from the accuracy of the whole data. {\color{black}We see that CML provides faster convergence than FLHB for both scenarios since it has access to the whole dataset at once, hence, the weight update becomes easier from that of FLHB.	The diversity of users' dataset increases in proportion to $K$, thus, the FLHB convergence rate decreases as $K$ increases. Moreover,  the training performance of both CML and FLHB is slightly better in \emph{scenario 2} than \emph{scenario 1} since the former provides better data representation with finer angular sampling for user locations, whereas the latter includes i.i.d. data, which has slightly poorer representation. In prior works, FL is proven to have significant performance loss for non-i.i.d. data as compared to i.i.d. scenario~\cite{FL_Bennis2,FL_Bennis3,FL_Bennis5,FL_traffic,elbir2020federated,FL_Gunduz}. Hence, it is worth noting that our experimental setup for \emph{scenario 2} is different than that of conventional non-i.i.d. datasets so that they can provide more realistic user distribution scenario. For example, if non-i.i.d. dataset was used for \emph{scenario 2}, then, we could observe significant changes in the angular location of the users, which is physically not possible. Thus, we model the dataset in \emph{scenario 2} with users located in non-overlapping angular sectors. By doing so, we have introduced diversity in the local datasets of users.}
	
	

	\begin{figure}[t]
		\centering
		{\includegraphics[draft=false,width=\columnwidth]{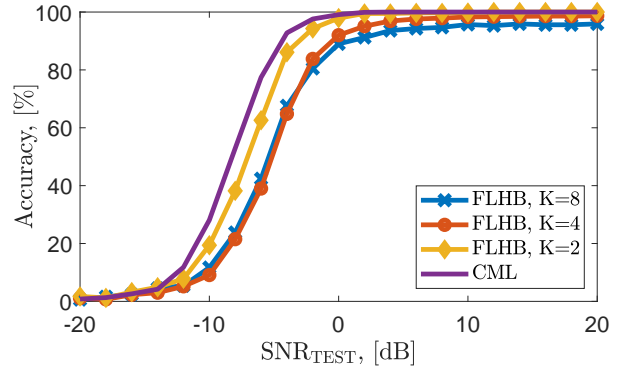} } 
		\caption{Validation accuracy with respect to $\mathrm{SNR}_\mathrm{TEST}$ for \emph{scenario 2}.
		}
		\label{fig_ValACC}
	\end{figure}

	\begin{figure}[t]
		\centering
		{\includegraphics[draft=false,width=\columnwidth]{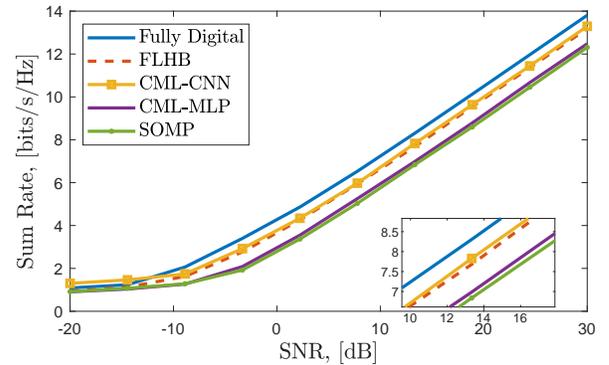} } 
		\caption{Spectral efficiency versus $\mathrm{SNR}$.
		}
		\label{fig_Rate}
	\end{figure}
	In Fig.~\ref{fig_ValACC}, we present the accuracy of the corrupted channel data for \emph{scenario 2}. The level of channel corruption is represented by $\mathrm{SNR}_\mathrm{TEST}$, defined similar to $\mathrm{SNR}_\mathrm{TRAIN}$. {\color{black}CML has  better performance than FLHB. Moreover, the performance of FLHB deteriorates as $K$ increases due to the corruptions in the model aggregation caused by the data diversity.}
	
	Fig.~\ref{fig_Rate} shows the sum-rate for the competing techniques when  $\mathrm{SNR}_\mathrm{TEST}= 5$ dB. {\color{black} FLHB outperforms other algorithms including the CML-based MLP technique, since FLHB leverages the CLs providing better feature representation. Using the same NN structures, CML-CNN has slightly better performance than FLHB since CML-CNN has access the whole dataset at once whereas FLHB employs decentralized training.}

\begin{figure}[t]
		\centering
		{\includegraphics[draft=false,width=\columnwidth]{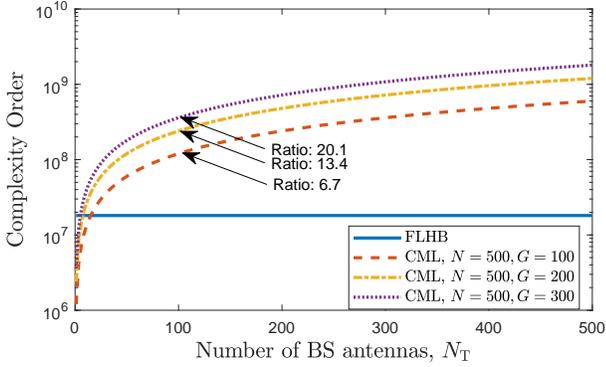} } 
		\caption{Transmission overhead complexity with respect to $N_\mathrm{T}$.
		}
		\label{fig_Complexity}
	\end{figure}

	{\color{black}While the above results present incremental performance losses for FLHB over CML, the main advantage of FLHB is on the transmission overhead. Hence, we compare the overhead of the transmission of the data/model parameters for the proposed approach FLHB and CML in Fig.~\ref{fig_Complexity}.} The transmission overhead of {\color{black}FLHB} is fixed as the total number of learnable parameters of CNN, which is given by $30\cdot 603648 =18109440$ where we assume training takes $30$ iterations. On the other hand, {\color{black}CML} has the transmission overhead due to the size of the dataset which is $\textsf{D} =3\cdot N\cdot G\cdot K $.  As it is seen, {\color{black}FLHB} provides fixed and lower transmission overhead than {\color{black}CML}. In particular, {\color{black}FLHB} is more efficient than {\color{black}CML} on the order of magnitude of  $\{6.7,13.4, 20.1\}$ for $G = \{100, 200, 300\}$, respectively.
	
	{\color{black}We also compare the time complexity of the competing algorithms to estimate the hybrid beamformers. Given the channel matrix as input, MLP and SOMP require approximately  $56$ and $320$ milliseconds ($ms$) for $N_\mathrm{T}=100$. Since FLHB and CML-CNN use the same NN architecture, they have the same time complexity of approximately $53$ $ms$ while FLHB has much lower training overhead, as shown in Fig.~\ref{fig_Complexity}.  }
	
\begin{figure}[t]
		\centering
		{\includegraphics[draft=false,width=\columnwidth]{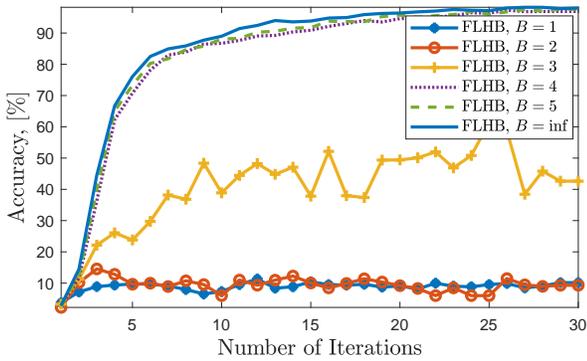} } 
		\vspace*{-6mm}
		\caption{Training performance with model parameter quantization.
		}
		\label{fig_Quantization}
	\end{figure}

	{\color{black}Finally, Fig.~\ref{fig_Quantization} shows  the effect of model parameter quantization. The model weights in each layer are quantized separately as in~\cite{elbirQuantizedCNN2019} before transmitting to the BS. We see that at least $B=4$ bits are required for accurate model training. 
	}
	
	\section{Conclusions}
	\label{sec:Conc}
	In this work, we propose a federated learning  strategy for hybrid beamforming ({\color{black}FLHB}) for mm-Wave massive MIMO systems. {\color{black}FLHB} is advantageous since it does not require the whole  training dataset to be sent to the BS for model training, instead, only the gradient information is used to update the NN weights. {\color{black}FLHB} provides more robust and tolerant performance than {\color{black}CML} against the imperfections in the input data.	While {\color{black}FLHB} exhibits superior performance as compared to the conventional {\color{black}CML} training, the following issues can be studied for future considerations. 1) While FL reduces the overhead of the transmitted data length, the gradient information can further be reduced via compression techniques. 2) The scheduling time of the users for training should be minimized because training requires the collection of the data of all users. {\color{black} 3) Asynchronous FL strategies are also of great interest where the users are scheduled with ``first come, first served'' fashion by a master node.}

	\bibliographystyle{IEEEtran}
	\footnotesize{\bibliography{IEEEabrv,references_062_journal}}
	\balance

\end{document}